\documentclass[nofootinbib,a4paper,aps,pra,showpacs,preprintnumbers,twocolumn]{revtex4-1}
\usepackage{cmap}
\usepackage{graphics,graphicx,epsfig}
\usepackage{epstopdf}
\usepackage[centertags]{amsmath}
\usepackage{amsfonts}
\usepackage{amssymb}
\usepackage{amsthm,color}
\usepackage{euscript}
\DeclareGraphicsExtensions{.pdf,.png,.jpg,.eps}
\usepackage[T1]{fontenc}
\usepackage[utf8]{inputenc}
\usepackage{xcolor}
\usepackage{hyperref}
\definecolor{linkcolor}{HTML}{000000} 
\definecolor{urlcolor}{HTML}{000000} 
\hypersetup{pdfstartview=FitH, linkcolor=linkcolor,urlcolor=urlcolor, colorlinks=true}

\begin{document}
  
\title{Teleportation with a cubic phase gate}
\author{E. R. Zinatullin}
\author{S. B. Korolev}
\author{T. Yu. Golubeva}
\affiliation{St. Petersburg State University, Universitetskaya nab. 7/9, St. Petersburg, 199034, Russia}
\begin{abstract}
We propose a modified quantum teleportation scheme to increase the teleportation accuracy by applying a cubic phase gate to the displaced squeezed state. We have described the proposed scheme in Heisenberg's language, evaluating it from the point of view of adding an error in teleportation, and have shown that it allows achieving less error than the original scheme. Repeating the description in the language of wave functions, we have found the range of the displacement values, at which our conclusions will be valid. Using the example of teleportation of the vacuum state, we have shown that the scheme allows one to achieve high fidelity values.
\end{abstract}
\maketitle
  

\section{Introduction}

Quantum teleportation is one of the basic protocols of quantum information processing \cite{Bennett, Vaidman, Bouwmeester, Braunstein, Furusawa}. It is this protocol that underlies one of the promising models of universal quantum computation - one-way quantum computation model \cite{Menicucci, Raussendorf, Nielsen}. In our work, we will discuss the continuous-variable quantum teleportation protocol \cite{Lloyd, Braunstein2}. Unlike discrete quantum systems, the use of continuous-variable ones allows one to build deterministic schemes. However, working with continuous-variable quantum systems also has a significant drawback: the presence of unavoidable errors associated with the finite squeezing degree of states, which are used as a resource for teleportation. Continuous-variable one-way quantum computation has inherited this disadvantage. The squeezing, which is experimentally achievable at the moment, turns out to be insufficient for performing universal quantum computations: the maximum experimentally achievable squeezing degree is -15 dB \cite{Vahlbruch}, whereas for universal computations (without postselection procedure) the squeezing of -20.5 dB \cite{Menicucci1} is required. In this regard, the reduction of unavoidable errors for continuous-variable quantum computation remains an important theoretical problem.

One option to improve the accuracy of teleportation is to use entangled non-Gaussian states as the main resource. This effect was experimentally demonstrated in \cite{Opatrny, Cochrane}. To obtain such a resource, one uses the procedure of conditional subtraction or addition of photons applied to the Gaussian entangled state. This procedure is probabilistic, which means that teleportation protocol using such non-Gaussian states will no longer be deterministic. We asked ourselves: what other non-Gaussian operations can be used to increase the accuracy of teleportation? Can we get any new benefits from using them? Is it possible to increase the accuracy of teleportation while remaining within the framework of deterministic processes?

As the transformation under study, we have chosen the cubic phase transformation. The first idea of cubic phase states generation was proposed by Gottesman, Kitaev, and Preskill back in 2001 \cite{GKP, Ghose, Gu}. It turned out that this idea is difficult to implement in practice since it requires performing the quadrature displacement operation by a value far from what is achievable in an experiment. Because of this, the cubic phase gate has long remained just an abstract mathematical transformation. However, the situation has changed in recent years. There are more and more works devoted to new methods for the cubic phase states generation \cite{Yukawa, Yanagimoto, Hillmann, YZhang} and the implementation of the cubic phase gate \cite{Miyata}. As a result, the cubic phase gate gradually turns from a purely theoretical transformation into a real-life device.

In the article \cite {Zinatullin} we showed that it is possible to improve the accuracy of teleportation while remaining within the framework of Gaussian transformations. By replacing the two beamsplitters with CZ transforms, we were able to reduce the error in one of the quadratures by using the weight coefficients of the CZ transformations \cite{Larsen, Su2018, Alexander}. In this article, we modify the teleportation scheme we proposed earlier, using a cubic phase gate to reduce the error level in another quadrature. In section \ref{s1}, we describe the teleportation procedure in Heisenberg's language and evaluate the quality of teleportation in terms of adding a teleportation error. Then, in the next section, we will once again describe the teleportation procedure, but this time in the language of wave functions without any approximations, estimate the fidelity of the teleported state and demonstrate the limits of applicability of the proposed approach and its practical feasibility.


\section{Teleportation with a cubic phase gate in Heisenberg's language} \label{s1}

\begin{figure}[t]
\begin{center}
\includegraphics[width=85mm]{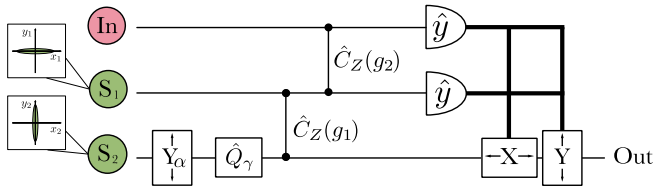}
\caption{Teleportation scheme using a cubic phase gate. In the diagram: $\text{In}$ is the input (teleported) state; $\text{S}_1$ and $\text{S}_2$ are oscillators squeezed in orthogonal quadratures; $\text{Y}_\alpha$ denotes the displacement of the $y$-quadrature by a fixed value $\alpha$; $\hat Q_\gamma$ is a cubic phase gate with a conversion coefficient $\gamma$; $\hat{C}_{Z}(g_i) $ is CZ transformations with weight coefficients $g_i$; $\hat y$ - homodyne detectors measuring the $y$-quadrature of the field in the channel; X and Y denote devices that displace the corresponding quadratures of the fields in the channel, depending on the detection results.}
\label{shem}
\end{center}
\end{figure}

Let us start with a description of the proposed teleportation scheme shown in Fig. \ref{shem}, in Heisenberg's language.
Two oscillators at the input of the scheme, designated in the figure as $ S_1 $ and $ S_2 $, are squeezed in orthogonal directions and are described by the following quadrature components: 
\begin{align}
\hat x_{s,1}&=e^r \hat x_{0,1}, \qquad \hat y_{s,1}=e^{-r} \hat y_{0,1},\\
\hat x_{s,2}&=e^{-r} \hat x_{0,2}, \qquad \hat y_{s,2}=e^r \hat y_{0,2},
\end{align}
where $\hat x_{0, j} $ and $ \hat y_{0, j} $ are the quadratures of the $j$ oscillator in the vacuum state. We are performing a series of manipulations on the second oscillator. First, we need to displace its $y$-quadrature.  The displacement operator, which displaces the $y$-quadrature of the $j$-th oscillator by the classical quantities $\alpha$, has the following form:
\begin{align}
\hat Y_{\alpha,j}=e^{i \alpha \hat x_j}. \label{d_op}
\end{align}
We displace the $y$-quadrature of the second oscillator by the classical fixed value $\alpha$ (transformation $\text {Y}_ \alpha$)
\begin{align}
\hat a_2'=\hat x_{s,2}+i(\alpha+\hat y_{s,2}). \label{disp}
\end{align}
For further reasoning, we need to impose the following condition on the value of $\alpha$: $\alpha^2 \gg \langle \hat y_{s, 2}^2 \rangle $. 

\begin{figure}[t]
\begin{center}
\includegraphics[width=85mm]{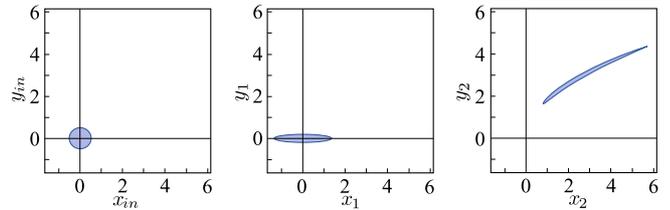}
\caption{Schematic uncertainty regions of the input state and resource states after a cubic phase gate.}
\label{Obl1}
\end{center}
\end{figure}

In the second step, we apply a cubic phase gate ($\hat Q_\gamma $) to this oscillator. The action of this gate on the j-th oscillator is defined by the following operator:
\begin{align}
\hat Q_{\gamma,j}=e^{-i\gamma \hat y_j^3}, \label{qfg}
\end{align}
where $\gamma$ is the coefficient of cubic phase transformation. After applying the cubic phase gate, the oscillator in the second channel will proceed to the non-Gaussian state, which is described by the Eq. 
\begin{align}
\hat a_2''=\hat x_{s,2}+3\gamma(\alpha+\hat y_{s,2})^2+i(\alpha+\hat y_{s,2}).
\end{align}
It should be noted that, due to the displacement by $\alpha$, the quadrature values of the second oscillator lie in the first quadrant of the phase plane. A schematic representation of the uncertainty regions for each of the three oscillators in this cut of the scheme is shown in Fig. \ref {Obl1}. Note that here and below, as an illustrative example of the scheme operation, we chose the input (teleported) state as a vacuum one.

\begin{figure}[t]
\begin{center}
\includegraphics[width=60mm]{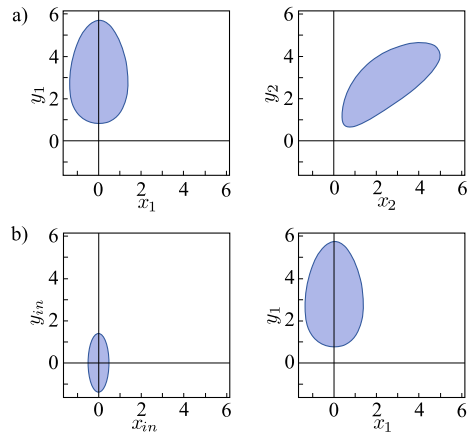}
\caption{ a) Schematic uncertainty regions of oscillators in the first and second channels after the first CZ transformation; b) schematic uncertainty regions of oscillators in the input and first channels after the second CZ transformation.}
\label{Obl2}
\end{center}
\end{figure}

After this, the states in the first and second channels are entangled using the CZ transformation, the operator of which has the following form: 
\begin{align}
\hat C_{z,jk}=e^{i g_{jk} \hat x_j \hat x_k}. \label{cz}
\end{align}
Here, the coefficient $g_{jk}$ is the so-called weight coefficient of the transformation, it can take any real values \cite{Larsen, Su2018, Alexander}. The weight coefficient of the first transformation CZ is denoted by $g_1$, then the amplitudes of the oscillators after the transformation will take the form:
\begin{align}
\hat a_1'&=\hat x_{s,1}+i\left(\hat y_{s,1}+g_1\left(\hat x_{s,2}+3\gamma(\alpha+\hat y_{s,2})^2\right)\right), \label{1cz_1}\\
\hat a_2'''&=\hat x_{s,2}+3\gamma(\alpha+\hat y_{s,2})^2+i(\alpha+\hat y_{s,2}+g_1 \hat x_{s,1}). \label{1cz_2}
\end{align}
After this transformation, the uncertainty region of the oscillator in the second channel is stretched along the $y$-quadrature but remains in the first quadrant of the phase plane (see Fig. \ref{Obl2} (a)). At the same time, the uncertainty region of the oscillator in the first channel will be strongly stretched in the $y$-quadrature and will displace to the upper half-plane.

Next, using the second CZ transform with the weight coefficient $ g_2 $, we entangle the state in the first channel with the input state we want to teleport. As a result, we get:
\begin{align}
&\hat a_{in}'=\hat x_{in}+i (\hat y_{in}+g_2 \hat x_{s,1}), \label{2cz_1}\\
&\hat a_1''=\hat x_{s,1}+i\big(\hat y_{s,1}+g_1\left(\hat x_{s,2}+3\gamma(\alpha+\hat y_{s,2})^2\right) \nonumber\\
&\qquad+g_2 \hat x_{in}\big), \label{2cz_2}\\
&\hat a_2'''=\hat x_{s,2}+3\gamma(\alpha+\hat y_{s,2})^2+i(\alpha+\hat y_{s,2}+g_1 \hat x_{s,1}).
\end{align}
As shown in Fig. \ref{Obl2} (b), after the second CZ transformation, the uncertainty region of the input state will be stretched along the $y$-quadrature, and the state uncertainty region in the second channel will practically not change. After that, we measure the stretched $y$-quadratures of the input and first oscillators. The corresponding operators of photocurrents will be equal to
\begin{align}
&\hat i_{y,in}=\beta (\hat y_{in}+g_2 \hat x_{s,1}),\\
&\hat i_{y,1}=\beta \big(\hat y_{s,1}+g_1\left(\hat x_{s,2}+3\gamma(\alpha+\hat y_{s,2})^2\right) \nonumber\\
&\qquad\;\;+g_2 \hat x_{in}\big), \label{i_y1}
\end{align}
where $\beta$ is the amplitude of the homodyne detector's  local oscillator. As a result of a single measurement, we obtain some values of the photocurrents $i_ {y, in}$ and $i_{y, 1}$, and the measured values of the quadratures will be $y_ {in, m}=i_ {y, in}/\beta$ and $y_{in, m}=i_{y,in}/\beta$ and $y_{1,m}=i_{y, 1}/\beta$, respectively.

Such a measurement, due to the entanglement of the resource state, will lead to a change in the quadrature components of the field in the second channel:
\begin{align}
\hat x_2'''=&-\frac{g_2}{g_1}\hat x_{in}-\frac{\hat y_{s,1}}{g_1}+\frac{y_{1,m}}{g_1}, \label{x2'''}\\
\hat y_2'''=&-\frac{g_1}{g_2}\hat y_{in}+\frac{g_1 \, y_{in,m}}{g_2}+\frac{1}{\sqrt{3\gamma}} \nonumber\\
&\times\sqrt{\frac{y_{1,m}}{g_1}-\frac{g_2}{g_1}\hat x_{in}-\frac{\hat y_{s,1}}{g_1}-\hat x_{s,2}}. \label{y2'''}
\end{align}
Here the operators of photocurrents are replaced by c-numbers corresponding to the results of specific measurements. Since the value of $\alpha$ is large enough, ambiguity disappears in the Eq. (\ref{y2'''}), and it is sufficient to take into account only positive values of the square root. In addition, again by choosing the value $\alpha$ one can ensure the correctness of the condition $y_{1,m}^2 \gg g_2^2 \langle \hat x_{in}^2 \rangle+\langle \hat y_ {s, 1 }^2 \rangle+ g_1^2 \langle \hat x_{s, 2}^2 \rangle$. If for some reason it is difficult to increase $\alpha$ so that the specified requirement is met, then one can use the procedure of postselection of measurement results discarding $ y_ {1, m}$ that do not satisfy the required condition.
Then we can expand the square root in the Eq. (\ref{y2'''}) in a series, keeping only the terms of the first order of smallness:
\begin{align}
\hat y_2'''=&-\frac{g_1}{g_2}\hat y_{in}+\frac{g_1 \, y_{in,m}}{g_2}+\frac{1}{\sqrt{3\gamma}} \bigg(\sqrt{\frac{y_{1,m}}{g_1}} \nonumber\\
&-\frac{g_2 \hat x_{in}}{2\sqrt{g_1 y_{1,m}}}-\frac{\hat y_{s,1}}{2\sqrt{g_1 y_{1,m}}}-\frac{\sqrt{g_1}\hat x_{s,2}}{2\sqrt{y_{1,m}}}\bigg). \label{y2'''_1}
\end{align}

For the transformation described by the Eqs. (\ref {x2'''}) and (\ref{y2'''_1}) to be a teleportation transformation, it is necessary to set $g_1=-g_2 \equiv g$. To complete the teleportation, it is necessary to displace the $x$-quadrature of the second oscillator by $-y_{1,m}/g$, and the $y$-quadrature by $y_{in, m}-\sqrt{y_ {1 , m}/(3 \gamma g)}$. Operators are defined similarly to Eq. (\ref{d_op}). Then the states at the output of the scheme will have the following form
\begin{align}
&\hat x_{out}=\hat x_{in}-\frac{\hat y_{s,1}}{g},\\
&\hat y_{out}=\hat y_{in}+\frac{1}{2\sqrt{3\gamma \, y_{1,m}}} \left(\sqrt{g}\hat x_{in}-\frac{\hat y_{s,1}}{\sqrt{g}}-\sqrt{g}\hat x_{s,2}\right).
\end{align}
In the resulting expressions, the first terms correspond to the desired teleportation effect. The rest of the terms are the noise added to the teleported state during the transformation. Let us estimate the magnitude of this noise and compare the result with the traditional teleportation scheme.
It is convenient to characterize the error level by the magnitude of the mean-square errors of each quadrature, $\langle \delta \hat{e}^2_x \rangle= \langle (\hat x_{out} - \hat x_{in})^2\rangle$, $\langle \delta \hat{e}^2_y \rangle= \langle (\hat y_{out} - \hat y_{in})^2\rangle$. In our case, they will have the following form:
\begin{align}
&\langle \delta \hat{e}^2_x \rangle=\frac{1}{g^2} \langle\hat y_{s,1}^2\rangle, \label{er_x}\\
&\langle \delta \hat{e}^2_y \rangle=\frac{1}{12\gamma \, y_{1,m}} \left(g \langle\hat x_{in}^2\rangle+\frac{1}{g}\langle\hat y_{s,1}^2\rangle+g\langle\hat x_{s,2}^2\rangle\right). \label{e_y}
\end{align}
To estimate the magnitude of the error, we will replace in the Eq. (\ref{e_y}) the measured value of the quadrature $y_{1,m}$ by its  average value, which according to the Eq. (\ref{i_y1}) will be equal to $\langle\hat y_1\rangle=\langle\hat i_{y,1}/\beta\rangle \approx 3g\gamma\alpha^2$. The main error in the $y$-quadrature is introduced by the term with $\hat x_{in}$, since the noise in the $\hat y_{s, 1}$ and $\hat x_{s, 2}$ quadratures of the resource oscillators is suppressed. Thus, the error in $y$-quadrature can be estimated as 
\begin{align}
\langle \delta \hat{e}^2_y \rangle\approx\frac{1}{36\gamma^2 \alpha^2}\langle\hat x_{in}^2\rangle.
\label{er_y}
\end{align}
From the Eq. (\ref{er_x}), we see that the error suppression in the $x$-quadrature of the teleported state is proportional to the weighting coefficient of the CZ transformations. As mentioned above, this coefficient can theoretically take on arbitrarily large values. The possibilities of real manipulation of the weight coefficient in the teleportation scheme are studied in detail in the work \cite{Zinatullin}. The Eq. (\ref{er_y}) shows that the error in $y$-quadrature can be suppressed by increasing the displacement of $\alpha$.

There are two important facts to note:
\begin{enumerate}
\item We have a limitation on the ability to noise suppression due to non-Gaussianity. It is associated with a physical restriction on the maximum value by which we can perform the displacement operation. However, if vacuum or noise suppressed states are used as the teleportable states (which is a typical situation for applications), this restriction will not play a significant role.
\item We can independently reduce the error in one of the quadratures due to the weight coefficients of the CZ transformation, and in the other due to the non-Gaussian resource. Also,  owing to the selection based on the measurement results, it is possible to set the lower limit of the teleportation accuracy.
\end{enumerate}

\begin{figure}[t]
\begin{center}
\includegraphics[width=85mm]{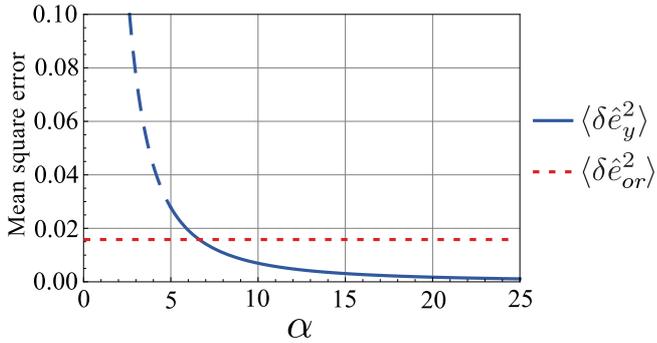}
\caption{The value of the mean-square error of the teleported state $y$-quadrature depending on the displacement magnitude $\alpha$ of the non-Gaussian resource. The red dotted line marks the level of teleportation error in the original scheme.}
\label{err}
\end{center}
\end{figure}

Fig. \ref{err} shows the dependence of the magnitude of the mean-square error of the $y$-quadrature on the displacement value $\alpha$. For calculations, the value of the conversion coefficient of the cubic phase was taken as $\gamma =0.1$ \cite {Miyata}. The red dotted line denotes the mean-square error for the original teleportation scheme $\langle \delta \hat{e}^2_{or} \rangle=2e^{-2r}\langle \delta \hat{e}^2_{vac} \rangle$ at the maximum currently experimentally achievable squeezing degree of -15 dB \cite{Vahlbruch}. As we can see from the graph, for values of $\alpha> 7$, our proposed scheme will outperform the classical teleportation scheme. Let us pay attention to the fact that for small $\alpha$ our approximations will be incorrect, therefore the dependence is shown by a dashed line. In this range of values, the displacements are too small, and the quadrature values of the second oscillator will not completely lie in the first quadrant of the phase plane. This entails ambiguity in the Eq. (\ref{y2'''}) and the emergence of a state like the Schr{\"o}dinger's cat state \cite{Sokolov}, which in our case leads to a rapid increase in the error variances.

Let us see what energy is required to carry out the displacement by the value $\alpha=20$. For a numerical estimate, we take the laser wavelength $\lambda=430$ nm and the intensity transmittance of the asymmetric beam splitter $\tau = 0.01 $ (the parameters are taken from the article \cite{Miwa}). The amplitude of the laser field required to implement such a displacement is given by the following expression:
\begin{align}
E=\sqrt{\frac{\hbar\omega}{2\varepsilon_0 V}} \, \frac{\alpha}{\sqrt{\tau}},
\end{align}
where $\omega$ is the laser frequency, and $V$ is the volume occupied by the laser pulse. The energy density of the electromagnetic wave $w=\varepsilon_0|E|^2$, then the energy of the laser pulse should be equal to
\begin{align}
W=&\int_V \varepsilon_0 |E|^2 dv=\frac{h c \alpha^2}{2\lambda \tau}=\frac{6.63 \cdot 10^{-34} \cdot 3 \cdot 10^8 \cdot 20^2}{2 \cdot 430 \cdot 10^{-9} \cdot 0.01}\nonumber\\
\approx &10^{-14} \, \text{J}.
\end{align}
The required laser power turns out to be relatively small, so the procedure described by us is feasible. Note that the existing idea of the experimental complexity of the displacement procedure refers to a significantly different range of values of $\alpha$. For comparison, when generating a cubic phase state by the method described in the articles \cite{Ghose, Gu}, displacement of the order of $\alpha=10^{12}$ are required. Under similar conditions, this would require pulse energy of the order of $ 10^7 $ J, which, of course, is far from the conditions of a modern quantum-optical experiment.

\section{Description of the teleportation procedure in the language of wave functions} \label{s2}

Heisenberg's approach, which we used in the previous section, is illustrative and makes it easy to assess the quality of teleportation. However, it has several disadvantages. First, it does not allow us to directly assess the quality of teleportation of specific quantum states (to obtain expressions in terms of fidelity). In addition, we had to resort to approximations that limit the applicability of our result. It is also worth mentioning that Heisenberg's approach may give incorrect results when describing transformations over non-Gaussian resources \cite{Sokolov}. Bearing in mind all of the above, we repeat the above procedure but use the language of wave functions to describe it without resorting to any approximations.

It will be convenient for us to decompose the oscillators vector-states in terms of the eigenstates of the $x$-quadrature operator -- $\{|x\rangle\}$. In this case, the vector-state of some input state $|\psi_{in}\rangle$ and squeezed resource oscillators $|\psi_1\rangle$ and $|\psi_2\rangle$ can be represented as
\begin{align}
&|\psi_{in}\rangle=\int dx_{in} \, \psi_{in}(x_{in}) |x_{in}\rangle,\\
&|\psi_1\rangle=\int dx_1 \, \psi_s(x_1;-r) |x_1\rangle,\\
&|\psi_2\rangle=\int dx_2 \, \psi_s(x_2;r) |x_2\rangle.
\end{align}
Here, for convenience, we have introduced the notation for the wave function of the squeezed state with the squeezing coefficient $r$:
\begin{align}
\psi_s(x;r)=\sqrt[4]{\frac{e^{2r}}{\pi}}\exp\left(-\frac{e^{2r} x^2}{2}\right).
\end{align}
In doing so, the eigenstates of $x$ and $y$-quadratures are related to each other as
\begin{align}
&|x\rangle=\frac{1}{\sqrt{2\pi}}\int dy \, e^{-ixy} |y\rangle, \\
&|y\rangle=\frac{1}{\sqrt{2\pi}}\int dx \, e^{ixy} |x\rangle.
\end{align}

Let us repeat the operations that we performed on the states in the previous section. First, similar to the Eq. (\ref{disp}), we displace the $y$-quadrature of the second oscillator by the classical value $\alpha$
\begin{align}
|\psi_2'\rangle&=e^{i\alpha \hat x_2}|\psi_2\rangle=\int dx_2 \, e^{i\alpha x_2} \psi_s(x_2;r) |x_2\rangle. \label{disp2}
\end{align}
Here we have used the fact that $|x_2 \rangle$ are eigenstates of the displacement operator $e^{i\alpha \hat x_2}$. Therefore, in the Eq. (\ref{disp2}) under the integral sign, we can replace the operator $\hat x_2$ with the corresponding eigenvalue $x_2$. In what follows, we will use similar reasoning, replacing operators with numbers. If in Eq. (\ref{disp2}) we proceed to the decomposition in terms of eigenstates of the $y$-quadrature operator, then the action of the displacement operator will correspond to the shift of the oscillator wave function by value $\alpha$:
\begin{align}
|\psi_2'\rangle=\int dy_2 \, \psi_s(y_2-\alpha;-r) |y_2\rangle.
\end{align}
After that, we apply the cubic phase operation to this oscillator, the action of which is described by the Eq. (\ref{qfg}). Then we get that
\begin{align}
|\psi_2''\rangle=e^{-i\gamma \hat y_2^3}|\psi_2'\rangle=\int dy_2 \, e^{-i\gamma y_2^3} \psi_s(y_2-\alpha;-r) |y_2\rangle.
\end{align}
Let us go back to the eigenstate decomposition of the $x$-quadrature operator. We can represent the vector-state of the oscillator in the second channel as
\begin{align}
|\psi_2''\rangle=\int dx \, \psi_{2}''(x) |x\rangle,
\end{align}
here its wave function is given by the expression
\begin{align}
\psi_{2}''(x)=\frac{1}{\sqrt{2\pi}} \int dy_2 \, e^{i\gamma y_2^3} e^{i x y_2} \psi_s(y_2-\alpha;-r).
\end{align}

Then we sequentially apply two transformations CZ (Eq. (\ref {cz})) with weight coefficients $g_1$ and $g_2$. The first CZ transformation entangles resource oscillators (see (\ref{1cz_1})-(\ref{1cz_2})), and the second entangles the input state with the state in the first channel (see (\ref {2cz_1})-(\ref{2cz_2})). In addition, from the last section, we know that the transformation we are describing will be a teleportation only under the condition $g_1= -g_2 \equiv g$. Then after the first transformation CZ we get that
\begin{align}
|\psi_1',&\psi_2'''\rangle=e^{ig \hat x_1 \hat x}|\psi_1\rangle\otimes|\psi_2''\rangle\nonumber\\
&=\iint dx_1 dx \, e^{ig x_1 x} \psi_s(x_1;-r) \psi_2''(x) |x_1\rangle\otimes|x\rangle,
\end{align}
and the vector-state of the system after the second CZ transformation will have the form:
\begin{align}
|\psi_{in}',\psi_1''&,\psi_2'''\rangle=e^{-ig \hat x_{in} \hat x_1}|in\rangle\otimes|\psi_1',\psi_2'''\rangle \nonumber\\
=&\iiint dx_{in} dx_1 dx \, e^{ig x_1 (x- x_{in})} \psi_s(x_1;-r) \psi_{2}''(x) \nonumber\\
&\times \psi_{in}(x_{in}) |x_{in}\rangle\otimes|x_1\rangle\otimes|x\rangle.
\end{align}

Next, we measure the $y$-quadratures of the input and first oscillators. In this case, the measured values of quadratures are equal to $y_{in,m}$ and $y_{1,m}$, respectively. Homodyne measurement is an operation of projecting onto the eigenstates of $y$-quadrature, which correspond to the measured values of the photocurrents, i.e., onto state
\begin{align}
|y_{in,m}\rangle \otimes |y_{1,m}\rangle=&\frac{1}{2\pi}\iint dx_{in}' dx_1' \, e^{ix_{in}'y_{in,m}} e^{ix_1' y_{1,m}} \nonumber\\
&\times|x_{in}'\rangle \otimes |x_1'\rangle.
\end{align}
After the measurement, the non-normalized state of the second oscillator has the form
\begin{align}
|\psi_2'''\rangle=&\langle y_{in,m}| \otimes \langle y_{1,m}|\psi_{in}',\psi_1'',\psi_2'''\rangle\nonumber\\
=&\frac{1}{\sqrt{2\pi}}\iint  dx_{in} dx \, e^{-ix_{in}y_{in,m}} \psi_{2}''(x) \psi_{in}(x_{in}) \nonumber\\
&\times \psi_s\big(g(x-x_{in}-y_{1,m}/g);r\big) \, |x\rangle.
\end{align}
The resulting state must be normalized to the root of the probability density that when measuring the quadratures $\hat y_1$ and $\hat y_{in}$, one will obtain the values $y_{1,m}$ and $y_{in,m}$:
\begin{align}
P(y_{1,m},y_{in,m})=\langle\psi_2'''|\psi_2'''\rangle.
\end{align}

\begin{figure*}[t]
\begin{center}
\includegraphics[width=175mm]{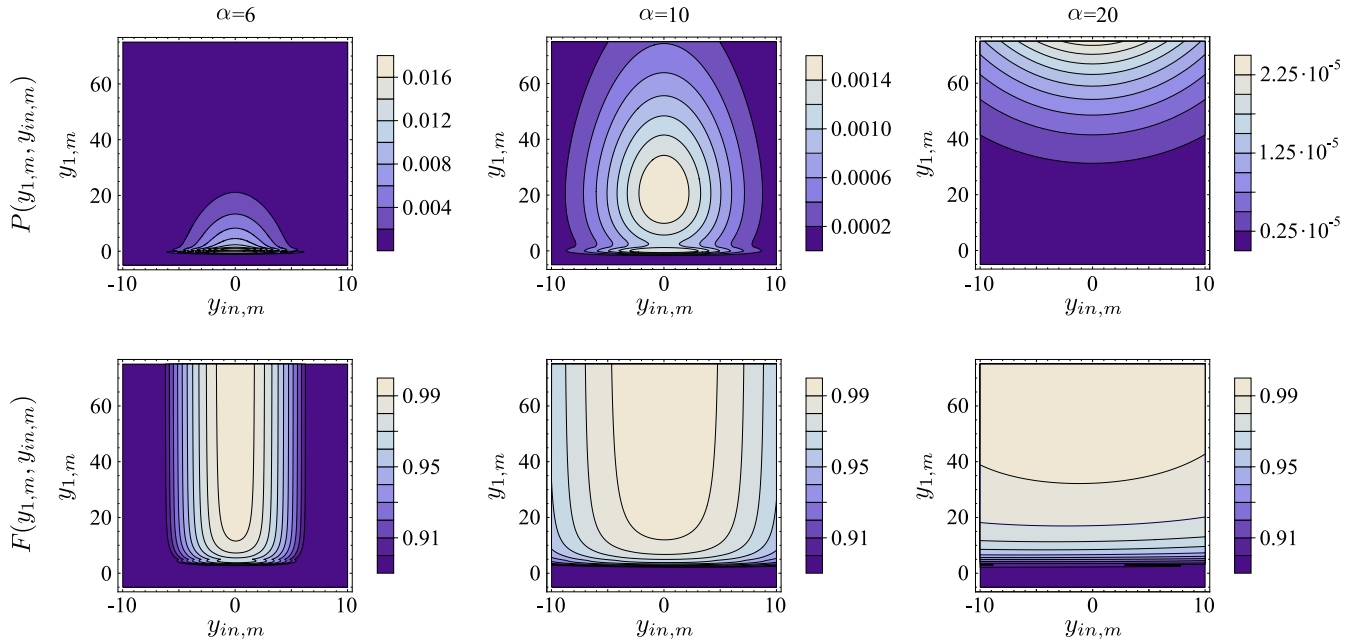}
\caption{The probability density of measuring the values of $y_ {1, m}$, $y_{in, m} $ and the fidelity value depending on $y_{1,m}$, $y_{in, m}$ during teleportation of the vacuum state for displacement values of the resource non-Gaussian state $\alpha=6,\, 10,\, 20$.}
\label{PF}
\end{center}
\end{figure*}

As we know from the previous section, to complete the teleportation procedure, it is necessary to displace the $x$-quadrature of the second oscillator by the value of $- y_{1, m}/ g$, and the $y$-quadrature by the value of $y_{in, m} - \sqrt{y_{1, m}/(3 \gamma g)}$. Thus, the vector of the teleported state will have the following form:
\begin{align}
|\psi_{out}\rangle=&e^{i \frac{y_{1,m}}{g} \hat y}e^{i (y_{in,m}-\sqrt{y_{1,m}/3\gamma g}) \hat x}|\psi_2'''\rangle\nonumber\\
=&\frac{1}{\sqrt{2\pi P(y_{1,m},y_{in,m})}}\iint  dx_{in} dx \, e^{-ix_{in}y_{in,m}} \nonumber\\
&\times \exp\left(i \left(y_{in,m}-\sqrt{\frac{y_{1,m}}{3\gamma g}}\right) \left(x+\frac{y_{1,m}}{g}\right)\right) \nonumber\\
&\times \psi_s\big(g(x-x_{in});r\big) \psi_{2}''\left(x+\frac{y_{1,m}}{g}\right) \psi_{in}(x_{in}) \, |x\rangle.
\end{align}

\begin{figure}[t]
\begin{center}
\includegraphics[width=60mm]{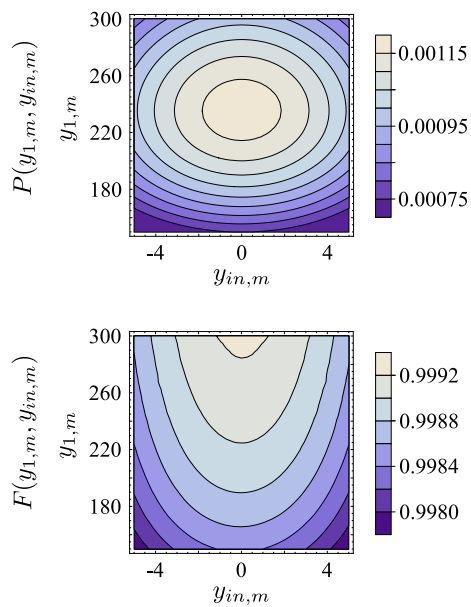}
\caption{The probability density of measuring the values of $y_ {1, m}$, $y_{in, m} $ and the fidelity value depending on $y_{1,m}$, $y_{in, m}$ during teleportation of the vacuum state for $\alpha=20$.}
\label{F20}
\end{center}
\end{figure}

With this approach to protocol analysis, we have the opportunity to assess the fidelity of teleportation. It is defined as
\begin{align}
F=|\langle \psi_{out}|\psi_{in}\rangle|^2.
\end{align}
In the calculations, as in the previous section, we took the vacuum state as the teleported state, and we set the squeezing degree of the resource oscillators equal to $-15$ dB. When teleporting, we will be interested in the case when the errors in both quadratures are approximately the same. According to the Eqs. (\ref{er_x}) and (\ref{er_y}), this is achieved when the weight coefficient of the CZ transformation is $ g = 6 \gamma \alpha e^ {-r}$. We calculated the probability density $P(y_ {1,m}, y_{in,m})$ of measuring the values $y_{1,m}$ and $y_{in, m}$ and the fidelity value $F(y_{1,m},y_{in, m}) $ for displacement value of $ \alpha=6, \,10,\,20$ (see Fig. \ref{PF}). In the first case, the displacement value is too small, and the approximations we used in section \ref {s1} will not be performed. The graphs show that the probability of successful teleportation with a value of $\alpha=6 $ is extremely small. The non-monotonic behavior of the graph in the vicinity of the bottom of the figure indicates the ambiguity of the values $y_2$ or the used resource state (the displacement of the quadrature is not large enough to transfer the entire uncertainty area to the first quadrant of the phase plane).  With $\alpha = 10 $ we can already teleport with $F>0.99$. However, this will require additional selection based on the measurement results, since approximately in 10 \% of cases, we will get the value $ y_ {1, m} <5$. If we increase the displacement value to $ \alpha=20$, the approximations from section \ref{s1} will be fully fulfilled, and we will no longer need additional selection based on the measurement results. As you can see from Fig. \ref{F20}, fidelity values above 0.998 are achieved in the most probable range.


\section{Conclusion}

In the presented work, we have shown that it is possible to reduce the teleportation error in one of the quadratures using a cubic phase gate. We have demonstrated this by analyzing our scheme in terms of adding a teleportation error in Heisenberg's language. In addition, we have described the scheme in the language of wave functions and have demonstrated that to fulfill the approximations made by us, it is necessary to perform relatively small displacements of the squeezed state before applying the cubic phase transformation. These displacements can be implemented in practice. If necessary, the average teleportation accuracy can be increased by selection based on the measurement results. This selection turns out to be necessary if it is not possible to displace the squeezed state by a sufficient value.

It is worth noting that the scheme allows for some variations.
For example, by moving the displacement transformations and the cubic phase gate from the second channel to the first one and swapping the states of the oscillators at the input of the first and second channels, one can achieve an additional reduction in the error in the $x$-quadrature. In this case, the error reduction will occur simultaneously due to the weight coefficients of the CZ transformation and due to the cubic phase gate.

Compared to the photon subtraction teleportation scheme, our scheme has a significant advantage. With a sufficient displacement value, it works deterministically and does not require additional selection based on the measurement results. Also, we have an additional parameter that we can control -- the magnitude of the displacement of the squeezed state, on which the average error in one of the quadratures depends. It is worth noting that a natural drawback of the proposed scheme is its technical complication. Its main element is a cubic phase gate, the practical implementation of which is still an urgent task. In addition, the protocol involves the use of entanglement CZ gates, the practical implementation of which is noticeably more difficult than mixing fields at a beam splitter (see, for example, \cite{Zinatullin}).

Nevertheless, bearing in mind the active development of Gaussian and non-Gaussian resources, it can be argued that with the advent of more advanced methods for implementing a cubic phase gate, the proposed protocol can give a significant gain in comparison with the original teleportation scheme.

\vspace{0.5 cm}

This work was financially supported by the Russian Foundation for Basic Research (Grant No. 19-02-00204a). S. B. Korolev acknowledge a financial support from the Theoretical Physics and Mathematics Advancement Foundation “BASIS” (Grant No. 21-1-4-39-1).



 \end{document}